\documentclass[runningheads]{llncs}
\usepackage{graphicx}
\usepackage{tcolorbox}
\usepackage{booktabs}
\usepackage{float}
\usepackage{amsmath}
\graphicspath{ {images/} }

\begin{document}
\title{A Supervised Learning Approach For Heading Detection}

\author{\IEEEauthorblockN{}
\IEEEauthorblockA{\\
\\
Email: }
\and
\IEEEauthorblockN{}
\IEEEauthorblockA{Lakehead University\\
955 Oliver Rd, Thunder Bay, ON P7B 5E1\\
Email: }
}

\author{Sahib Singh Budhiraja \and
Vijay Mago \\ \email{(sbudhira,vmago)@lakeheadu.ca} }

\authorrunning{S. Budhiraja and V. Mago}
\titlerunning{A Supervised Learning Approach For Heading Detection}
%
\institute{
Lakehead University, 955 Oliver Rd, Thunder Bay, ON P7B 5E1 
}

\maketitle              

\begin{abstract}
As the Portable Document Format (PDF) file format increases in popularity, research in analysing its structure for text extraction and analysis is necessary. Detecting headings can be a crucial component of classifying and extracting meaningful data. This research involves training a supervised learning model to detect headings with features carefully selected through recursive feature elimination. The best performing classifier had an accuracy of 96.95\%, sensitivity of 0.986 and a specificity of 0.953. This research into heading detection contributes to the field of PDF based text extraction and can be applied to the automation of large scale PDF text analysis in a variety of professional and policy based contexts.\\

\end{abstract}

\section{Introduction}

As the amount of information stored within PDF documents increases worldwide, the opportunities for large scale text based analysis requires increasingly automated processes, as the amount of document processing is time consuming and labour intensive for human professionals.  Systematic processing and extraction of textual structure is increasingly necessary and useful as demonstrated in El-Haj et al.'s work involving 1500 financial statements\cite{ElHajUKCorpus}.  Categorizing data into seperate sections is quite easy for humans, as they rely on visual cues such as headings to process textual information. Machines, despite being able to process large amounts information at high speeds, require effort to classify and interpret text based data. This paper explores the application of supervised classifiers to operationalize a system that would aid in the identification of headings. PDF documents are a visually exact digital copy that displays text by drawing characters on a specific location \cite{pdfref} and present a challenge in analysis because the files do not provide enough information on how the text is organized and formatted. \\
A supervised classifier that is trained on labelled data provides one solution to categorizing PDF text as it tells the classifier how to make predictions based on the data provided. This research involved comparing and systematically testing a variety of classifiers for the purpose of selecting classifiers best suited to this application. \textit{Recursive feature elimination}\cite{GuyonRFE}is used to ensure the classifiers only use the best and minimum number of features for making predictions. Cross validation is used to tune the hyper parameters of a given machine learning algorithm for increased performance before testing it out on test data. The final trained classifier is currently being applied to detect headings in course outline documents and extract learning outcomes. The extracted learning outcomes are being used for automating the process of developing university/college transfer credit agreements by using semantic similarity algorithms\cite{loaga}.\\

\section{Related Work}
While PDF format is convenient as it preserves the structure of a document across platforms, extracting textual layout information is required for detecting headings and further analysis. One solution to extracting layout information is to convert the PDF into HTML and use the HTML tags for further analysis. Once converted to HTML all the information related to text formatting required for the analysis, like font size and boldness of text, can be easily extracted. A variety of PDF to HTML document tools are available and have been assessed based on the text and structural loss associated with each tool\cite{GoslinHTML}. Additional work includes PDF to HTML text detection approaches that maintain layout and font information\cite{JiangText}, table detection, extraction and annotation\cite{ShahExtraction} and analysis using white spaces\cite{RahmanLayout}. HTML conversion is clearly a well established approach to analyzing PDF layout and content.\\

\noindent Previous research provides insight into processes related to extracting the heading layout of a HTML document\cite{ManabeHeading}. In Manabe's work, headings are used to divide a document at certain locations that indicate a change in topic. Document Object Model(DOM) trees are used to sort candidate headings based on their significance and to define blocks. A recursive approach is applied for document segmentation using the list of candidate headings and evaluate with good results using a manually labelled dataset. \\

\noindent El-Haj et al. provide a practical application of document structure detection through the analysis of a large corpus of UK financial reports including 1500 annual reports from 200 different organizations. A list of \lq gold standard' section names was generated from 50 randomly selected reports and used to match with corresponding sections of every document page in the dataset. Section matches were then extracted and evaluated using sensitivity, specificity and F1 score in addition to being reviewed by a domain expert for accuracy\cite{ElHajUKCorpus}.\\

\noindent Current research has taken steps towards a system which analyses a document's textual structure. But there is a need to have an approach that can efficiently and accurately analyse the textual layout of a document and divide it into content sections to automate the process of extracting text from a PDF documents. We present out supervised learning approach for heading detection as a solution for it.\\

\section{Methodology}

\subsection{Data Collection}

Our data set consisted of 500 documents\footnote{Repository available at: https://github.com/sahib-s/Heading-Detection-PDF-Files} downloaded from Google using \textit{Google Custom Search API} \cite{gAPI}. To extract the correspoding formatting/style information the documents were converted from PDF to HTML using \textit{pdf2txt}, which is a PDFMiner wrapper available in Python \cite{pdfminer}. This is illustrated in Fig \ref{extraction} which shows some sample text and its corresponding HTML tags generated using the conversion process. The final data points are also shown in the Fig \ref{extraction}, which was generated by parsing the HTML tags using regular expressions. A regular expression is string of characters used to define a search pattern\cite{regex}. The regular expressions used for parsing the tags are as follows:\\

\begin{tcolorbox}

To extract Font size and corresponding text:\\

$r$\lq$<\setminus s*?span[^{\wedge}>]*font-size:(\setminus w*)px[^>]*>(.*?)<\setminus/span\setminus b[^>]*>$ \rq \\

To check if text is bold we look for the following regular expression for the word bold in the starting tag:\\

$r$\lq$[Bb]old$\rq

\end{tcolorbox}

\noindent Each data point contains some text, font size and a flag which is either 1 or 0 depending on the corresponding text being bold or not. The whole process yielded 83,194 data points, which was then exported into an Excel file for further pre-processing.

\subsection{Data Preprocessing}
The process of transforming raw data into usable training data is referred to as data preprocessing. The steps of data preprocessing for this research are as follows:


\begin{figure*}
\begin{center}

\includegraphics[width=12cm,height=18cm]{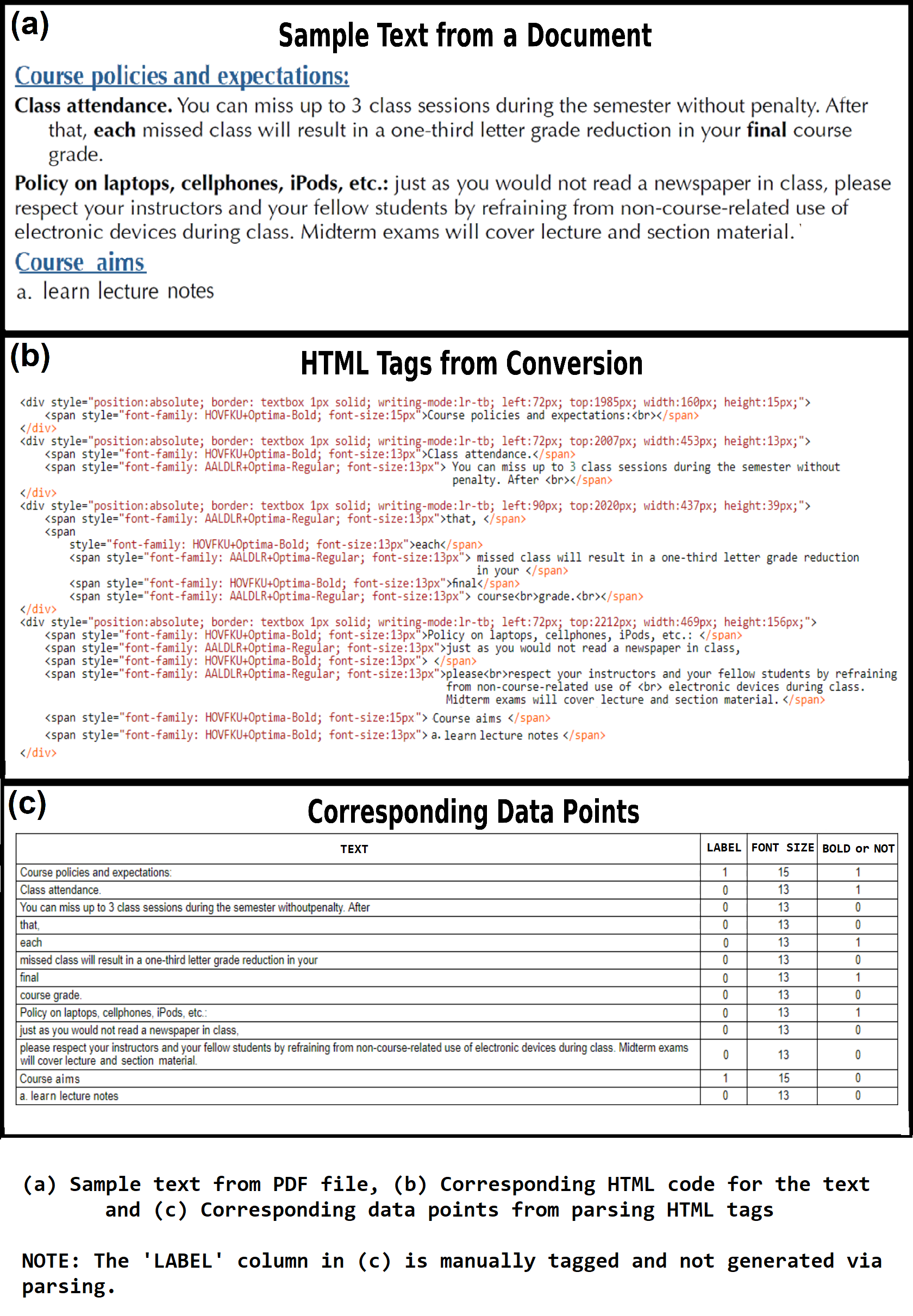}
\caption{Extraction of Data from Documents}
\label{extraction}       
\end{center}
\end{figure*}

\subsubsection{Data Labelling:}
Data labelling refers to the process of assigning data points labels, this makes the data suitable for training supervised machine learning models. All the 83914 data points are manually labelled by cross referring to the documents as both training and testing data needs to be labelled. If the text in the data point was a heading the label was set to 1 otherwise 0. Labelling data is one of the most important steps of preprocessing because the performance of the model depends on how well the data is labelled. Example of labelled data points is provided in Fig \ref{extraction}(c).

\subsubsection{Balancing The Dataset:}
The dataset is considered imbalanced if the prevalence of one class is more than the other. The number of headings in our dataset is very less as compared to non-headings, this is because of the fact that the number of headings in a document is far less than the number of other text elements. Sklearn's implementation for Synthetic Minority Over-sampling Technique (SMOTE) is used to balanced the dataset, which does so by creating synthetic data points for the minority class to make it even~\cite{sklearn,chawla2002smote}.

\subsubsection{Data Transformation:}
The process of transforming data into a form that has more predictive value is known as data transformation. The purpose of data transformation is to convert raw data points into \lq features' that contribute to more predictive value in training and decision making related to heading identification.  For example, font size and text are two features from the raw data which, in their base form, do not have much value but can be transformed into useful features for training an efficient model. The list of transformed data fields are as follows:

\begin{figure}
\begin{center}
\includegraphics[width=129mm,height=180mm]{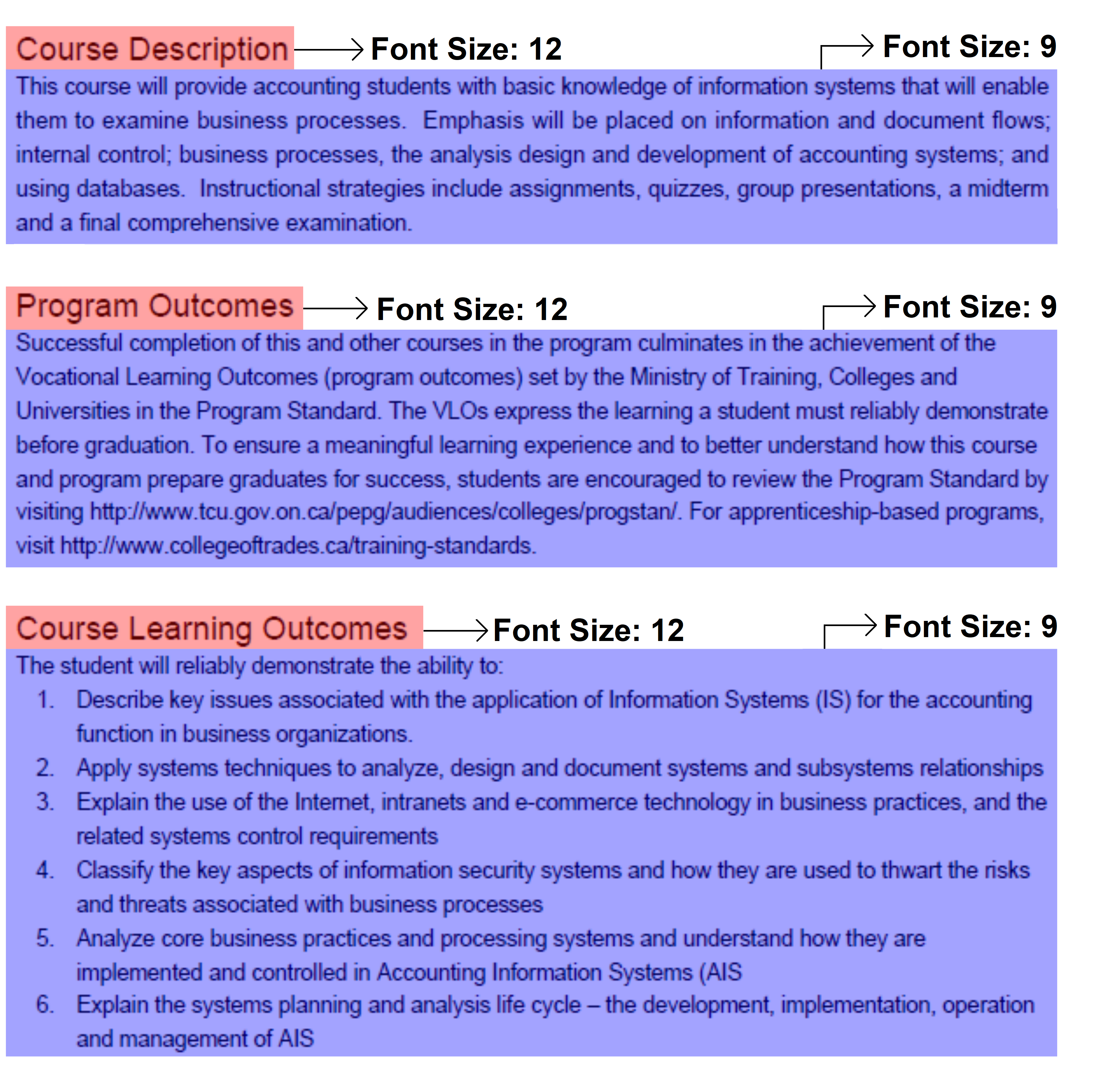}
\caption{Font Size Threshold Assumption Example}
\label{fontthresh}       
\end{center}
\end{figure}

\begin{itemize}
\item Font Flag: Headings tend to be larger in terms of font size as compared to the paragraph text that follows. Therefore, a higher font size increases the probability that the text is a heading. However, since each document is unique, there can not be a single threshold applied across all instances.\\
\noindent Thresholds are calculated for each document by measuring the frequency of each font size where each character with a particular font size is counted as one instance. The font size which has the maximum frequency is used as the threshold. This approach relies on the assumption that the most frequently used font size is the one that is being used for the paragraph text, so having any font size above that increases the probability of that text being a heading. Fig \ref{fontthresh} shows that the most frequently used font size is for the paragraph text with size 9 and all other text above it has more chances of being a heading. \textit{Font Flag} can take two possible values 0 and 1. If the font size for that data point is less than the corresponding  threshold then the value is set to 0, otherwise it is set as 1.
\item Text: The text is transformed into the following feature variables, which are also listed in Table \ref{fealist}.
\begin{itemize}
  \item Number of Words: The number of words in the text can be used for training, as headings tend to have less words when compared to regular sentences and paragraphs.
  \item Text Case: Headings mostly use title case, while sometimes they are in upper case as well. This variable tells whether the text is in upper case (all letters in upper case), lower case (all letters in lower case), title case (first letter of all words in uppercase) or sentence case (only the first letter of the text in uppercase).
  \item Features From Parts of speech(POS) Tagging: POS Tagging is the process of assigning parts of speech (verb, adverb, adjective, noun) to each word, which are referred to as tokens. The text from each data point is first tokenized and then each token is assigned a POS label~\cite{POSNLTK}. 
\end{itemize}
  
\noindent The POS frequencies provides the model with information on the grammatical aspect of the text and can be used to exploit the frequency of these labels in a text to identify headings and contribute to the accuracy of the model. For example, headings tend to have no verbs in them, though some might have them but absence of verbs increases the probability of the text being an heading. All frequency data collected from POS tagging is analysed in the feature selection process to differentiate between useful and irrelevant features collected through it. The frequency for each POS label is calculated and used to calculate the frequency of each POS tag in the text for each data point. These frequencies serve as potential features for the model. 
\end{itemize}
\noindent All these features brings the count of total number of features generated using the text to 11, 9 from POS tagging the text and 2 using its physical properties. \\
  
\begin{table}[H]
\centering
\caption{List of all features}
\label{fealist}
All features are integers, except for \textit{Bold or Not} and \textit{Font Threshold Flag} which are binary.
\vspace{1mm}
\begin{tabular}{|p{30mm}|p{90mm}|}
\hline
\textbf{Feature Name}               & \textbf{Description}                                                                                                                    \\ \hline
Characters                & Number of characters in the text.                                                                        \\
Words                     & Number of words in the text.                                                                             \\
Text Case                 & Assumes the value 0,1,2 or 3 depending on the text being being in lower case, upper case, title case or none of the three respectively. \\
Bold or Not               & Assumes the value 1 or 0 depending on the text being bold or not.                                                                       \\
Font Threshold Flag       & Assumes the value 1 or 0 depending on the font size of the text being greater than the threshold or not.              \\                  
Verbs                     & Number of verbs in the text.                                                                             \\
Nouns                     & Number of nouns in the text.                                                                             \\
Adjectives                & Number of adjectives in the text.                                                                        \\
Adverbs                   & Number of adverbs in the text.                                                                           \\
Pronouns                  & Number of pronouns in the text.                                                                          \\
Cardinal Numbers          & Number of cardinal numbers in the text.                                                                  \\
Coordinating Conjunctions & Number of coordinating conjunctions in the text.                                                         \\
Predeterminers            & Number of predeterminer  in the text.                                                                    \\
Interjections             & Number of Interjections in the text.                                                                     \\ \hline
\end{tabular}
\end{table}

\subsection{Feature Selection}
After pre-processing, 14 training features are established. There is a need to select the top features for building each individual model with maximum accuracy. Table \ref{fealist} lists all the features we are choosing from. To achieve this we used \textit{Recursive feature elimination} with Cross-Validation (RFECV), which recursively removes weak attributes/features and uses the model accuracy to identify features that are contributing towards increasing the predictive power of the model\cite{GuyonRFE}. The selection process is performed using the machine learning library, \lq\lq scikit-learn".\\

\noindent Cross validation is done by making 10 folds in the training set where one feature is removed per iteration. As per this analysis the accuracy does not increase on choosing to train the Decision Tree classifier with more than the following seven features:
\begin{itemize}
  \item Bold or Not
  \item Font Threshold Flag
  \item Number of words
  \item Text Case
  \item Verbs
  \item Nouns
  \item Cardinal Numbers
\end{itemize}

\noindent The same process is repeated for all the classifiers and their individual set of chosen features are listed in Table \ref{classfea}.

\subsection{Grid Search}

Tuning each classifiers parameters for optimal performance is performed using  accuracy from cross validation as a measure. We use various combinations of classifiers parameters and choose the combination with the best cross validation accuracy. This process is performed on various classifiers to choose their corresponding parameters. The description along with the final selected tuning parameters for each classifier used in this research are discussed in the next section.

\begin{table}[H]
\centering
\caption{Selected features for each classifier}
\label{classfea}
\vspace{6mm}
\begin{tabular}{|p{30mm}|p{90mm}|}
\hline
\textbf{Classifier Name}               & \textbf{Selected Features}                                                                                                                    \\ \hline
Decision Tree           		& Bold or Not, Font Threshold Flag, Words, Text Case, Verbs, Nouns, Cardinal Numbers  \\ & \\
SVM                     		& Bold or Not, Font Threshold Flag, Words, Text Case, Verbs, Nouns, Adjectives, Adverbs\\ & \\
k-Nearest Neaighbors    		& Bold or Not, Font Threshold Flag, Words, Verbs, Nouns, Adjectives, Cardinal Numbers, Coordinating Conjunctions \\ & \\
Random Forest           		& Bold or Not, Font Threshold Flag, Words, Text Case, Verbs, Nouns, Adverbs, Cardinal Numbers, Coordinating Conjunctions  \\ & \\
Gaussian Naive Bayes    		& Bold or Not, Font Threshold Flag, Words, Verbs, Nouns, Adjectives, Cardinal Numbers, Coordinating Conjunctions  \\     & \\      
Quadratic Discriminant Analysis & Bold or Not, Font Threshold Flag, Words, Verbs, Nouns, Adjectives, Coordinating Conjunctions  \\ & \\
Logistic Regression             & Bold or Not, Font Threshold Flag, Words, Text Case, Verbs, Nouns, Adverbs, Coordinating Conjunctions  \\ & \\
Gradient Boosting               & Bold or Not, Font Threshold Flag, Words, Text Case, Verbs, Nouns, Cardinal Numbers  \\ & \\
Neural Net                      & Bold or Not, Font Threshold Flag, Words, Text Case, Verbs, Nouns, Cardinal Numbers  \\
 \hline
\end{tabular}
\end{table}

\subsection{Training}

After the most suitable features and parameters for each classifier have been selected, we can proceed with training the classfiers using scikit-learn \cite{sklearn}.  

\subsubsection{Decision Tree}
Decision trees are the most widely used amongst classifiers as they have a simple flow-chart like structure starting from a root node. It branches off to further nodes and terminating at a leaf node. At each non-leaf node a decision is made, which selects the branch to follow. The process continues to the point where a leaf node is reached, which contains the corresponding decison\cite{decitree}. Gini impurity is used as a measure for quality of a split, which tells if the split made the dataset more pure.  Using Gini makes it computationally less expensive as compared to entropy which involves computation of logarithmic functions. The \lq\lq best" option for strategy chooses the best split at each node. The minimum number of samples required to split an internal node is set to 2 and the minimum number of samples needed to be at a leaf node is set to 3. The code snippet for training this classifier with the chosen parameters is given in Box 1\\

\begin{center} Box 1:  Code Snippet for Training Decision Tree Classifier\end{center}

\begin{tcolorbox}

treeclf = DecisionTreeClassifier(criterion = \lq gini\rq, splitter = \lq best\rq, min\_samples\_split = 2, min\_samples\_leaf = 3)\\
treeclf = treeclf.fit(traindata, truelabels)

\end{tcolorbox}
  
\subsubsection{Support Vector Machine (SVM)} 
It is a classifier that uses multi-dimensional hyperplanes to make classification. SVM also uses kernel functions to transform the data in sucha way that it is feasible for the hyperplane to effectively partition classes\cite{svmclass}. The kernel used is radial basis function(rbf), degree of the polynomial kernel function is set to 3 and gama is set to \lq\lq auto". The shrinking heuristics were enabled as they speed up the optimization. Tolerance for stop criteria is set to $2e-3$ and \lq ovr'(one vs rest) decision function is chosen for decision function shape. The code snippet for training this classifier with the chosen parameters is given in Box 2.\\

\begin{center} Box 2:  Code Snippet for Training Support Vector Machine Classifier\end{center}

\begin{tcolorbox}

svmclf = SVC(kernel=\lq rbf\rq, degree=3, gamma=\lq auto\rq, shrinking=True, tol=0.002, decision\_function\_shape=\lq ovr\rq)\\
svmclf = svmclf.fit(traindata, truelabels) 
 
\end{tcolorbox}

\subsubsection{k-Nearest Neighbors}  
The main idea behind k-Nearest Neighbors is that it takes into account the class of its neighbors to decide how to classify the data point under consideration. Each neighbor’s class is considered as their vote towards that class and the class with the most votes is assigned to that data point\cite{nearest}. The number of neighbours used to classify a point is set to 10. Each neighbours are weighed equally as weights is set to \lq distance\rq. Minkowsky distance function used as the distance metric. The code snippet for training this classifier with the chosen parameters is given in Box 3.\\

\begin{center} Box 3:  Code Snippet for Training k-Nearest Neighbors Classifier\end{center}

\begin{tcolorbox}

neighclf = KNeighborsClassifier(n\_neighbors = 10, weights = \lq distance\rq, metric = \lq minkowski\rq)\\
neighclf = neighclf.fit(traindata, truelabels)

\end{tcolorbox}

\subsubsection{Random Forest} 
This classifier works by choosing random data points from the training set and creating a set of decision tress. The final decision regarding the class is made by aggreggation of the outputs from all the trees\cite{forest}. The number of trees in the forest is set to 2 and \lq gini' is used as a measure for quality of a split. The maximum depth of trees is set to 5 and the maximum number of features to be considered while searching for the best split is se to \lq auto'. The minimum number of samples required to split an internal node is set to 2 and the minimum number of samples needed to be at a leaf node is set to 3. The number of parallel jobs to running for both fit and predict is set to 1. The code snippet for training this classifier with the chosen parameters is given in Box 4.\\

\begin{center} Box 4:  Code Snippet for Training Random Forest Classifier\end{center}

\begin{tcolorbox}

RandomForestClassifier(n\_estimators = 2, criterion = \lq gini\rq, max\_depth = 5, max\_features=\lq auto\rq, min\_samples\_split=2, min\_samples\_leaf=3, n\_jobs=1)\\
rndForstclf = rndForstclf.fit(traindata, truelabels)

\end{tcolorbox}

\subsubsection{Gaussian Naive Bayes}
This classifier works by using Bayesian theorem with assumption of strong independence between the predictors(features). It is very useful for large data sets as it is quite simple to build and has no complicated iterative parameters\cite{naivebiasclas}.  This classifier does not have much to set when it comes to configuring parameters. Prior probabilities of the classes is set to [0.5, 0.5] as the number of headings is less as compared to other text. The code snippet for training this classifier with the chosen parameters is given in Box 5.\\

\begin{center} Box 5:  Code Snippet for Training Gaussian Naive Bayes Classifier\end{center}

\begin{tcolorbox}
gaussianclf = GaussianNB(priors = [0.5, 0.5])\\
gaussianclf = gaussianclf.fit(traindata, truelabels)
\end{tcolorbox}

\subsubsection{Quadratic Discriminant Analysis}
It works under the assumption that the measurements for each class are normally distributed while not assuming the covariance to be identical for all the classes. Discriminant analysis is used to choose the best predictor variable(s) and is more flexible than linear models making it better for a variety of problems\cite{discrimana}. Prior probabilities of the classes is set to [0.5, 0.5] as the number of headings is far less as compared to other text. The threshold used for rank estimation is set to $1e-4$. The code snippet for training this classifier with the chosen parameters is given in Box 6.\\
\begin{center} Box 6:  Code Snippet for Training Quadratic Discriminant Analysis Classifier\end{center}

\begin{tcolorbox}
quadclf = QuadraticDiscriminantAnalysis(priors = [0.5, 0.5], tol = 0.0001)\\
quadclf = quadclf.fit(traindata, truelabels)
\end{tcolorbox}

\subsubsection{Logistic Regression}
It is a discriminative classifier, therefore it works by discriminating amongst the different possible values of the classes\cite{logreg}. Penalization method is set to l2. The tolerance for stopping criteria is set to $2e-4$. The parameter \lq fit\_intercept' is set to true adding a constant to the decision function. The optimization solver used is \lq liblinear' and the maximum number of iterations taken for the solvers is set to 50. Multiclass is set to \lq ovr' fitting a binary problem for each label. The number of CPU cores used for parallelizing over classes is set to 1. The code snippet for training this classifier with the chosen parameters is given in Box 7.\\
\begin{center} Box 7:  Code Snippet for Training Logistic Regression Classifier\end{center}

\begin{tcolorbox}
logisticRegr = LogisticRegression(penalty=l2, tol=0.0002, fit\_intercept = True, solver=\lq liblinear\rq, max\_iter=50, multi\_class = \lq ovr\rq , n\_jobs=1)\\
logisticRegr = logisticRegr.fit(traindata, truelabels)
\end{tcolorbox}

\subsubsection{Gradient Boosting}
This classification method uses an ensemble of weak prediction models in a stage wise manner. In each stage, a weak model is introduced to make up for the limitations of the existing weak models\cite{gradboosting}. The loss function to be optimized is set as \lq deviance' and learning rate is set to 0.1. The minimum number of samples required to split an internal node is set to 2, the minimum number of samples needed to be at a leaf node is set to 1 and maximum depth of the individual regression estimators set to 3. The number of boosting stages is set to 150 and the measure of quality of a split is set to  \lq friedman\_mse\rq. The code snippet for training this classifier with the chosen parameters is given in Box 8.\\

\begin{center} Box 8:  Code Snippet for Training Gradient Boosting Classifier\end{center}

\begin{tcolorbox}
grdbstcf = GradientBoostingClassifier(loss = \lq deviance\rq, learning\_rate = 0.1, min\_samples\_split = 2, min\_samples\_leaf = 1, max\_depth = 3, n\_estimators = 150, subsample = 1.0, criterion = \lq friedman\_mse\rq)\\
grdbstcf = grdbstcf.fit(traindata, labels)
\end{tcolorbox}

\subsubsection{Neural Net}
This classifier works by imitating the neural structure of the brain. One data point is processed at a time and the actual classification is compared to the classification made by the classifier. Any errors recorded in the classification process are looped back into algorithm to improve classification performance in future iterations\cite{mago2011neural,neural}. The classifier is configured to have one hidden layer with 100 units. The activation function used for the hidden layer is \lq tanh'. The solver used for weight optimization is \lq lbfgs'. The batch rate is set to \lq auto' and the initial learning rate is set to 0.001. The parameter \lq max\_iter' is set to 300, which for \lq adam' solver defines the number of epochs. Sample shuffle is set to true, which enables sample shuffling in each iteration. The exponential decay rates for estimates of the first and second moment vector is set to 0.9 and 0.999 respectively. The code snippet for training this classifier with the chosen parameters is given in Box 9.\\

\begin{center} Box 9:  Code Snippet for Training Neural Net Classifier\end{center}

\begin{tcolorbox}
nurlntclf = MLPRegressor(hidden\_layer\_sizes = (100, ), activation = \lq tanh\rq, solver = \lq lbfgs\rq, learning\_rate = \lq invscaling\rq, batch\_size = \lq auto\rq, learning\_rate\_init = 0.001, max\_iter = 300, shuffle = True, beta\_1 = 0.9, beta\_2 = 0.999)\\
nurlntclf = nurlntclf.fit(traindata, labels)

\end{tcolorbox}

\begin{table}[]
\begin{center}
\caption{Classifier Accuracy}
Highest Value For Each Measure is Bold
\vspace{3mm}
\label{accuracy}
\begin{tabular}{|p{30mm}|c|c|c|c|c|}
\hline

\textbf{Classifier}                                 & \textbf{Sensitivity} & \textbf{Specificity} & \textbf{Precision} & \textbf{F1 Score}        & \textbf{Accuracy}  \\ 
\hline
Decision Tree                   & 0.986          & \textbf{0.952} & \textbf{0.953} & \textbf{0.970} & \textbf{96.95 \%}\\
SVM                             & 0.991          & 0.930          & 0.934          & 0.961          & 96.06 \% \\ 
K-Nearest Neighbors             & 0.979          & 0.945          & 0.946          & 0.962          & 96.22 \% \\ 
Random Forest                   & 0.991          & 0.928          & 0.932          & 0.961          & 95.99 \% \\ 
Gaussian Naive Bayes            & 0.981          & 0.912          & 0.917          & 0.948          & 94.66 \% \\ 
Quadratic Discriminant Analysis & 0.982          & 0.912          & 0.918          & 0.949          & 94.76 \% \\ 
Logistic Regression             & 0.982          & 0.904          & 0.911          & 0.945          & 94.34 \% \\ 
Gradient Boosting               & 0.991          & 0.941 		  & 0.944          & 0.967          & 96.66 \% \\          
Neural Net                      & \textbf{0.992} & 0.941          & 0.944          & 0.967          & 96.68 \% \\                                                                                            

\hline
\end{tabular}
\end{center}
\end{table}

\section{Test Results}

\begin{figure*}
\begin{center}
\includegraphics[width=12cm,height=6cm]{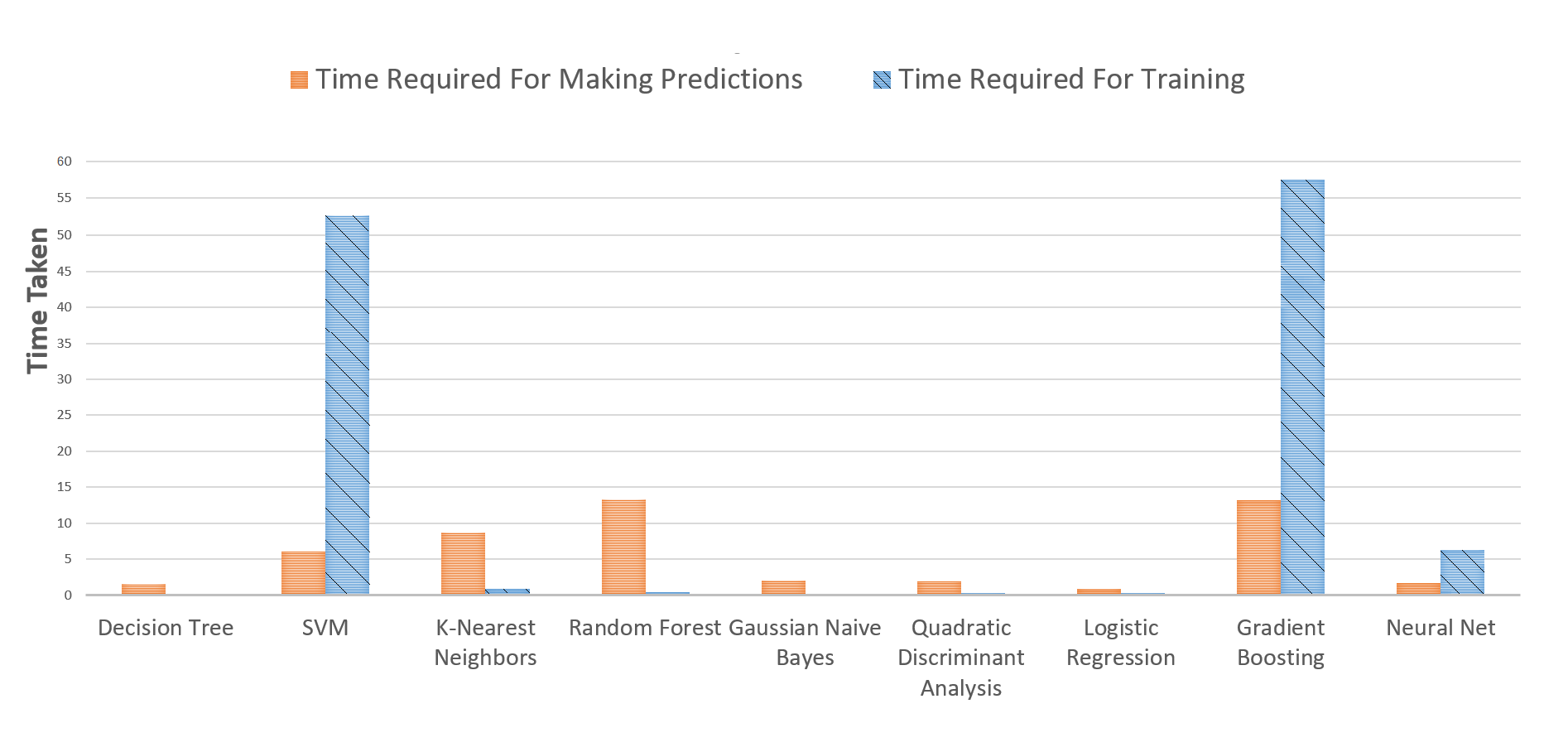}
\caption{Time required to train classifiers and run predictions on test data}
\label{results}       
\end{center}
\end{figure*}

\textbf{Training and Prediction Time:}
When dealing with a large number of documents, the time required to train a model and make predictions is important and is dependant on the type of classifier used, the number of features and the amount of data points. In this research all classifiers are trained using the same number of features and data points, therefore \lq time taken\rq provides a good measure of variations in training and prediction speed associated with each different classifier being used. Of note, the training time for a classifier should be considered in context, as training only needs to be performed once and can be saved for later use.  Therefore, a model that takes a long time to train can still be practical so long as it does not take a lot of time to make predictions. Fig \ref{results}. shows time required for training and making predictions using these classifiers. Time shown is average of 10 observations, which is done to reduce the effect of programs running in the background on the comparison.

\noindent \textbf{Confusion Matrix Based Evaluation:} We use evaluation parameters like Sensitivity, Specificity, Precision, F1 score and Net Accuracy calculated using confusion matrix to compare them to each other. Table \ref{accuracy} shows the results of this evaluation.\\
\textbf{ROC Curves and AUC:} A receiver operating characteristics(ROC) curve is used to visualize the trade-offs between sensitivity and specifity. These graphs are used for performance based selection of classifiers. The graph can be reduced to a numerical measure, AUC(or AUROC) which is the area under the ROC graph with values ranging from 0 to 1\cite{roc}. Table \ref{aucscores} shows the AUC scores for the classifiers used in this research. The discussion section provides more information on how we used AUC score to select the best classifier.

\begin{table}[H]
\centering
\caption{AUC Values for all Classifiers}
\vspace{3mm}
\label{aucscores}
\begin{tabular}{|c|c|}
\hline

\textbf{Classifier}                                 & \textbf{AUC}   \\ 
\hline
Decision Tree                              & \textbf{0.98}  \\
SVM                                        & 0.97  \\ 
K-Nearest Neighbors                        & 0.96  \\ 
Random Forest                              & 0.97  \\ 
Gaussian Naive Bayes                       & 0.96  \\ 
Quadratic Discriminant Analysis            & 0.96  \\ 
Logistic Regression                        & 0.95  \\ 
Gradient Boosting                          & \textbf{0.98}  \\          
Neural Net                                 & \textbf{0.98} \\                                                                                            

\hline
\end{tabular}
\end{table}

\section{Discussion}
\label{disc}
We recorded the time (in seconds) required for training each classifier and also time for making predictions as shown in Fig \ref{results}. Time taken by a classifier to make predictions is important when processing documents in bulk as it can increase the processing time. Time taken to train a classifier only has to be done once therefore it is not given that much importance. The Decision Tree Classifier took the least time for training while Gradient Boosting took the most. On comparing the prediction time Logistic Regression takes the least time and Random Forest takes the most. While prediction time is not the most important factor while choosing a classifier we take it into consideration when two classifiers are performing approximately the same.\\ 

\noindent The top three classifiers based on net accuracy are Decision Tree, Gradient Boosting, and Neural Network, however classifier selection can not solely rely on accuracy\cite{aucbetter,Lingaucbetter}. Therefore, we  also weigh the metrics like AUC, F1 score, sensitivity, and specificity to choose the best suited classifier for detecting headings.\\

\noindent The top three classifiers in terms of F1 score, precision, sensitivity and specificity are Decision Tree, Gradient Boosting, and Neural Network and the top 3 in terms of AUC as shown in Table \ref{aucscores} are again Decision Tree, Gradient Boosting, and Neural Network. The system is going to be dealing with documents in bulk and the prediction time for Decision Tree is better when compared to both Gradient Boosting and Neural Network. Therefore, we would be choosing our configuration of the Decision Tree for making the classifications.\\

\section{Testing The Generalizability}

Testing the chosen classifier on a general set of documents is important to show that it performs well on documents other than course outlines. We tested the chosen Decision Tree classifier on 12,919 data points collected from documents like reports and articles\footnote{Repository available at: https://github.com/sahib-s/Generalizability/}.  These data points were manually tagged using a survey. All the participants were graduate students from computer science department and were asked to point out headings and subheadings in the documents. Table \ref{gentest} shows the results which are equivalent if not better as compared to when tested on course outlines.\\

\begin{table}[H]
\centering
\caption{Test Results For General Set}
\label{gentest}
\begin{tabular}{|c|c|}
\hline

\textbf{Category}                                 & \textbf{Value}   \\ 
\hline
Total Data points          & 12919     \\
Sensitivity                & 0.928     \\ 
Specificity                & 0.966     \\ 
Precision                  & 0.964     \\ 
F1 SCORE                   & 0.946     \\ 
Accuracy                   & 94.73 \%  \\ 
AUC                        & 0.97      \\ 

\hline
\end{tabular}
\end{table}

\begin{table}[H]
\centering
\caption{Pearson Correlation Coefficient Between Each Feature Used in the Selected Classifier and Final Decision Labels}
\label{pear}
\vspace{2mm}
\begin{tabular}{|p{50mm}|c|}
\hline
\textbf{Feature Name}               & \textbf{Pearson Correlation Coefficient}                                                                                                                    \\ \hline
Bold or Not               & 0.7022 \\
Font Threshold Flag       & 0.2385 \\  
Words                     & 0.1389 \\
Verbs                     & 0.1229 \\
Nouns                     & 0.1207 \\
Cardinal Numbers          & 0.1201 \\
Text Case                 & 0.0660 \\

\hline
\end{tabular}
\end{table}

\section{Analysing The Results}
The discussed configuration of Decision Tree is best suited to detect heading as discussed in Section \ref{disc}. Analyzing the contribution of each feature towards the final decision made by the classifier is also important to understand the implications of the results. Table \ref{pear} shows the pearson correlation coefficient for all the features used in the selected classifier and final decision label. The list is in descending order of pearson correlation coefficient, therefore the top feature in the table contribute the most towards the final decision. Each feature was removed from the classifier one at a time and drop in evaluation metrics also verify the order of contribution presented by using the pearson correlation coefficient.  Therefore, the top three contributing features are the ones that rely on the physical attributes of the text.

\section{Extending The Classifier}

\noindent The extension of this work includes tagging of multiple labels like heading, paragraph text, header/footer text and table text. While classifying paragraph text is possible using the existing features, for properly classifying table text and header/footer text more data features are necessary. We are currently looking features from our white space detection approach discussed in chapter 5 and bounding box data from PDF to XML conversion to provide the model with what it needs to make this classification.  
 
\section{Conclusion}
This research has provided a structured methodology and systematic evaluation of a heading detection system for PDF documents. The detected headings provide information on how the text is structured in a document.  This structural information is used for extracting specific text from these documents based on the requirements of the field of application.  This supervised learning approach has demonstrated good results and we are currently applying our configuration of the Decision Tree classifier in the field of post-secondary curriculum analysis to identify headings and extract learning outcomes from course outlines for a research being conducted at DATALAB, Lakehead University, Canada.


\section*{Acknowledgment}

This research would not have been possible without the financial support provided by Ontario Council on Articulation and Transfer (ONCAT) through Project Number-2017-17-LU. We would also like to express our gratitude towards the datalab.science team and Andrew Heppner for their support.

\end{document}